\documentclass{PoS}

\usepackage[]{graphicx}
\usepackage{amsmath}
\usepackage{amssymb}

\title{Cosmological models with multiple dark matter species and long-range scalar interactions}

\ShortTitle{Multi-coupled Dark Energy cosmologies}

\author{\speaker{Marco Baldi}\\
        Excellence Cluster Universe,
        Boltzmannstrasse 2, 85748, Garching\\
        Dipartimento di Fisica e Astronomia, Universit\`a di Bologna,
        Viale B.~Pichat 6/2, 40127, Bologna\\
        E-mail: \email{marco.baldi5@unibo.it}}

\abstract{In this talk, I have discussed the implications of a multi-component nature of cosmic Dark Matter for the observational bounds on possible long-range fifth-forces mediated by a Dark Energy scalar field. By assuming a simple internal symmetry of the Dark Matter component associated to opposite coupling "charges" of two different particle species, the effects of Dark Energy interactions on both the background and linear perturbations evolution are strongly suppressed during the whole matter dominated phase, thereby relaxing present bounds on the coupling strength. The associated attractive and repulsive fifth-forces, however, might still have a very significant impact on the nonlinear dynamics of  collapsed structures. I have also described how some of these nonlinear effects are identified through dedicated cosmological N-body simulations as {\em i)} a possible fragmentation of bound Dark Matter halos into smaller objects, and {\em ii)} a consequent suppression of the nonlinear matter power at small scales. Both effects are potentially observable and might allow to further constrain the model.}

\FullConference{Proceedings of the Corfu Summer Institute 2012 "School and Workshops on Elementary Particle Physics and Gravity"\\
		September 8-27, 2012\\
		Corfu, Greece}

\begin{document}

\section{Introduction}

The overwhelming evidence for the Universe being presently dominated for about 95\% of its total energy budget by particles and fields that do not belong to the standard model of particle physics (see e.g. the recent results of the Planck satellite mission, \cite{Planck_016}) did not provide yet a sufficiently firm indication
about the fundamental nature of such cosmic "Dark Sector". Even its traditional split into two distinct and independent components identified with Dark Matter and Dark Energy has been recently challenged and is found to be mostly conventional in terms of model-independent observable quantities \cite{Amendola_etal_2013}. Therefore, although the minimal concordance scenario based on a cosmological constant and on a single family of collisionless Cold Dark Matter particles keeps to be proven fully consistent with data, a higher level of internal complexity of the dark sector cannot yet be excluded. Such additional complexity might involve either of the two conventionally-defined constituents of the dark sector, or possibly both, allowing a non-trivial dynamics and/or non-negligible perturbations and interactions of the Dark Energy field (see e.g. \cite{Wetterich_1988,Ratra_Peebles_1988,Creminelli_etal_2009,Wetterich_1995,Amendola_2000,Farrar2004}), as well as non-cold, non-collisionless, or multi-component nature of the Dark Matter fraction (see e.g. \cite{Khlopov_1995,Palazzo_etal_2007}). Even models where a single field is assumed to be responsible of the phenomenology of both Dark Energy and Dark Matter have been proposed and widely investigated \cite{Bertacca_etal_2007}; alternatively, a modification of gravity at large scales might be invoked as the source of the observed accelerated expansion of the Universe \cite{Hu_Sawicki_2007}. For a comprehensive review of possible extensions to the standard $\Lambda $CDM scenario, see e.g. \cite{Euclid_TWG}. 

A wide range of such alternative scenarios have been investigated in recent years down to the level of nonlinear structure formation making use of specifically-designed N-body algorithms, allowing to identify peculiar features arising only in the nonlinear regime (for an overview of cosmological simulations in non-standard Dark Energy cosmologies, see e.g. \cite{Baldi_2012b}).\\

In the present work, I will review the main features and the possible observational footprints of a particular class of such extended cosmologies, characterised by a higher level of complexity in both the Dark Energy and the Dark Matter sectors. Such model, known as the "Multi-coupled Dark Energy" scenario \cite{Baldi_2012a,Baldi_2012c}, is characterised by the existence of two distinct species of Cold Dark Matter particles, interacting with individual couplings to an evolving Dark Energy scalar field. The results presented in this paper will show how an internal complexity of the Dark Matter component might result in significantly relaxed constraints on possible interactions between Dark Energy and Dark Matter, with associated scalar fifth-force interactions of gravitational strength being not easily ruled out through standard background, linear and nonlinear observables.\\

The paper is organised as follows. In Section~\ref{mcde}, I will review the main equations describing Multi-coupled Dark Energy cosmologies, both for what concerns the background expansion history and the evolution of linear perturbations; in Section~\ref{nonlinear}, I will describe the first attempts to simulate
the formation and evolution of nonlinear structures in Multi-coupled Dark Energy cosmologies through dedicated N-body simulations, and I will discuss the first results obtained with such analysis. Finally, in Section~\ref{concl} I will summarise the results and draw my conclusions.

\section{Multi-coupled Dark Energy}
\label{mcde}

\subsection{Background}
I will consider flat cosmologies where the role of the Dark Energy (DE) responsible for the observed accelerated expansion of the Universe
is played by a classical Quintessence scalar field $\phi $ moving in a self-interaction potential $V(\phi )$. We include in our analysis also the Cold Dark Matter (CDM) and radiation components of the Universe, and for simplicity we neglect the presence of baryons. 
In the present work, without loss of generality we will restrict our discussion to the case of an exponential \cite{Lucchin_Matarrese_1984,Ferreira_Joyce_1998} self-interaction potential in the form:
\begin{equation}
V(\phi ) = Ae^{-\alpha \phi /M_{\rm Pl}}\,. 
\end{equation}

We consider the possibility of a species-dependent coupling between the DE scalar field and other forms of gravitating energy in the Universe,
following the original proposal of \cite{Damour_Gibbons_Gundlach_1990}, which is the basic argument for standard interacting DE
models \cite{Wetterich_1995,Amendola_2000}. Interacting DE scenarios, also known as "coupled DE" (cDE) cosmologies, have been 
thoroughly investigated in the past concerning their effects on the background \cite{Amendola_2000,Farrar2004,Koyama_etal_2009,CalderaCabral_2009}, linear perturbations \cite{Amendola_2004,Pettorino_Baccigalupi_2008,Valiviita_etal_2009}, and nonlinear structure formation 
\cite{Maccio_etal_2004,Baldi_etal_2010,Baldi_2011a,Li_Barrow_2011,CoDECS} evolution.
Observational constraints have allowed to significantly restrict the viable parameter space for standard cDE models 
\cite{Bean_etal_2008,LaVacca_etal_2009,Xia_2009,Baldi_Viel_2010}, and in particular have bound the 
coupling value to a few percent of the standard gravitational strength, ruling out with very high significance a possible scalar interaction 
of strength comparable to standard gravity.

In the present discussion, we will show how a higher level of internal complexity in the CDM sector might allow to evade such constraints
and reconcile a possible interaction of gravitational strength in the dark sector with the expected observational evolution of a standard $\Lambda $CDM cosmology. More specifically, we will assume that the CDM budget of the Universe is made by two different types of particles, with identical physical properties except for the sign
of their coupling to the DE scalar field $\phi $. 
Similar types of multiple Dark Matter models have already been considered in the literature
in the context of Warm Dark Matter cosmologies (see e.g. \cite{Palazzo_etal_2007,Maccio_etal_2012}) 
and also for the case of interacting DE scenarios (see e.g. \cite{Farrar2004,Huey_Wandelt_2006,Amendola_Baldi_Wetterich_2008,Brookfield_vandeBruck_Hall_2008,Brax_etal_2010}). The peculiarity of the model discussed here, however, is to avoid the introduction of additional free parameters as compared to standard cDE models with a single CDM particle species, by associating the DE-CDM interaction to a sort of {\em "charge"}
of CDM particles as a consequence of a new hidden symmetry in the CDM sector.

The background evolution of the Universe in the context of such Multi-coupled DE (McDE) scenario can be described by the following system of dynamic field equations:
\begin{eqnarray}
\label{klein_gordon}
\ddot{\phi } + 3H\dot{\phi } + \frac{dV}{d\phi } &=& +C \rho _{+} - C \rho _{-}\,, \\
\label{continuity_plus}
\dot{\rho }_{+} + 3H\rho _{+} &=& -C \dot{\phi }\rho _{+} \,, \\
\label{continuity_minus}
\dot{\rho }_{-} + 3H\rho _{-} &=& +C \dot{\phi }\rho _{-} \,, \\
\label{continuity_radiation}
\dot{\rho }_{r} + 4H\rho _{r} &=& 0\,, \\
\label{friedmann}
3H^{2} &=& \frac{1}{M_{{\rm Pl}}^{2}}\left( \rho _{r} + \rho _{+} + \rho _{-} + \rho _{\phi} \right)\,,
\end{eqnarray}
where an overdot represents a derivative with respect to the cosmic time $t$, the CDM density is given by $\rho _{\rm CDM} = \rho _{+}+\rho _{-}$,
$H\equiv \dot{a}/a$ is the Hubble function, and $M_{\rm Pl}\equiv 1/\sqrt{8\pi G}$ is the reduced Planck mass.
The dimensional coupling constant $C$ is defined as:
\begin{equation}
C\equiv \sqrt{\frac{2}{3}}\frac{1}{M_{\rm Pl}}\beta \,,
\end{equation}
with $\beta = {\rm const.} \geq 0$ being the usual definition (see e.g. \cite{Amendola_2004,Baldi_2011a,CoDECS}) of the
dimensionless coupling between DE and CDM.

The interaction terms in Eqs.~\ref{continuity_plus}-\ref{continuity_minus} imply a variation of the mass of CDM particles of each species ($+$ and $-$) as a consequence of the
evolution of the DE field, according to the equation:
\begin{equation}
\label{mass_variation}
\frac{d\ln \left[ M_{\pm}/M_{\rm Pl}\right] }{dt} = \mp C\dot{\phi }\,,
\end{equation}
where $M_{\pm }$ is the mass of a CDM particle of the positively ($+$) or negatively ($-$) coupled species. It is important to notice that 
the dynamics of the DE scalar field determines an opposite variation of the mass of particles of the two different CDM types associated to their opposite couplings. As a consequence of such different evolution, the relative abundance of the two CDM particle species changes in time. We
quantify such evolution with the dimensionless asymmetry parameter defined as
\begin{equation}
\label{eta}
\mu  \equiv \frac{\Omega _{+} - \Omega _{-}}{\Omega _{+} + \Omega _{-}}\,,
\end{equation}
with the fractional density parameters $\Omega _{i}$ defined in the usual way as:
\begin{equation}
\Omega _{i} \equiv \frac{\rho _{i}}{3H^{2}M_{\rm Pl}^{2}}\,.
\end{equation}

The dynamical evolution of such McDE model is equivalent (see \cite{Brookfield_vandeBruck_Hall_2008}) to that of a standard cDE model
with a single CDM species and with an effective coupling given by
\begin{equation}
\label{effective_coupling}
\beta _{\rm eff }= \beta \left( \frac{\Omega _{+}}{\Omega _{\rm CDM}} - \frac{\Omega _{-}}{\Omega _{\rm CDM}}\right) = \beta \mu  \,.
\end{equation}
Such equivalence already shows how the internal complexity of the CDM sector might provide an effective screening mechanism of the DE interaction in the background, since the effective coupling is dynamically suppressed whenever the two CDM species share the same relative abundance.

\begin{figure}[h]
\includegraphics[height=54mm]{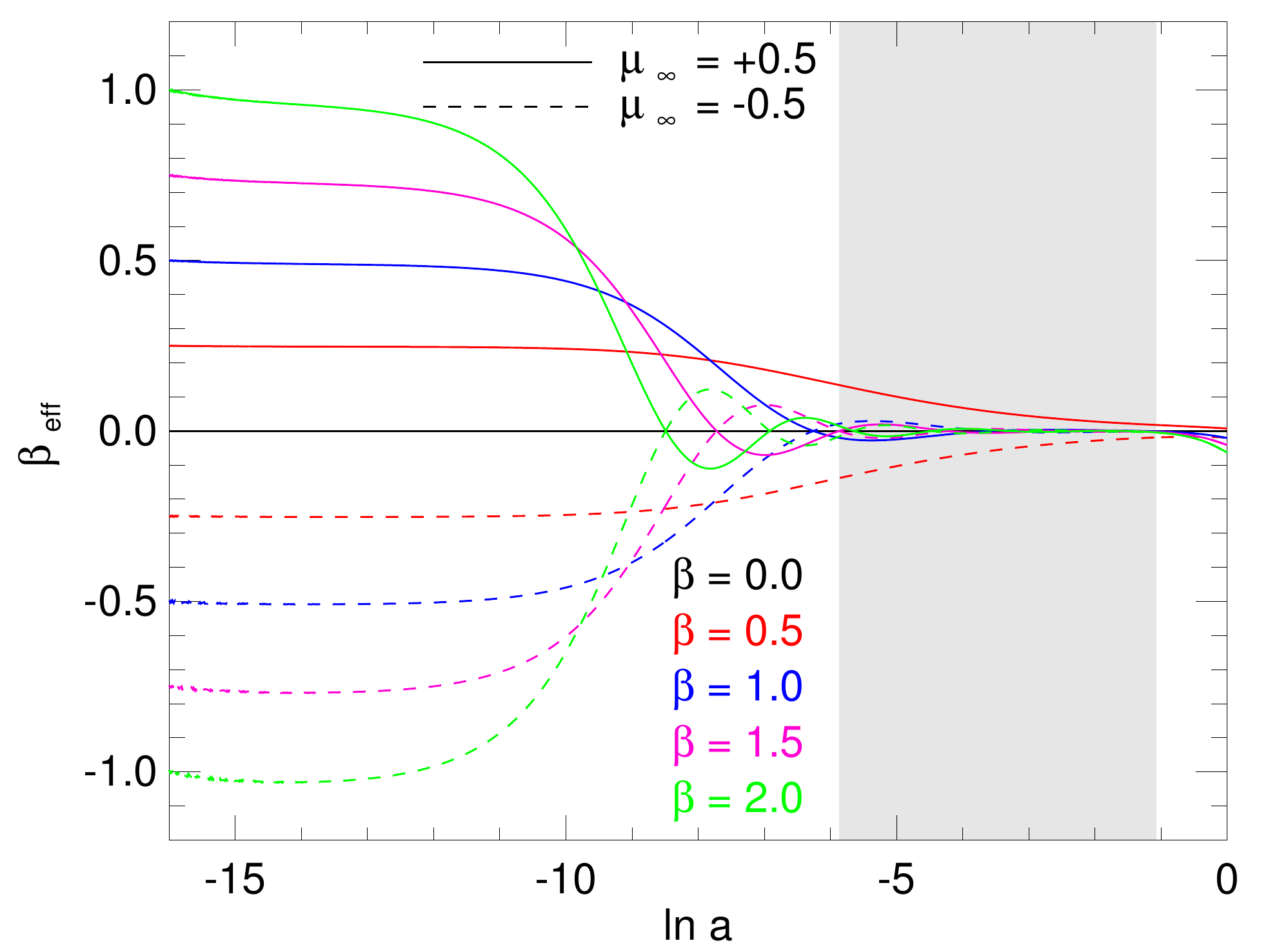}
\includegraphics[height=54mm]{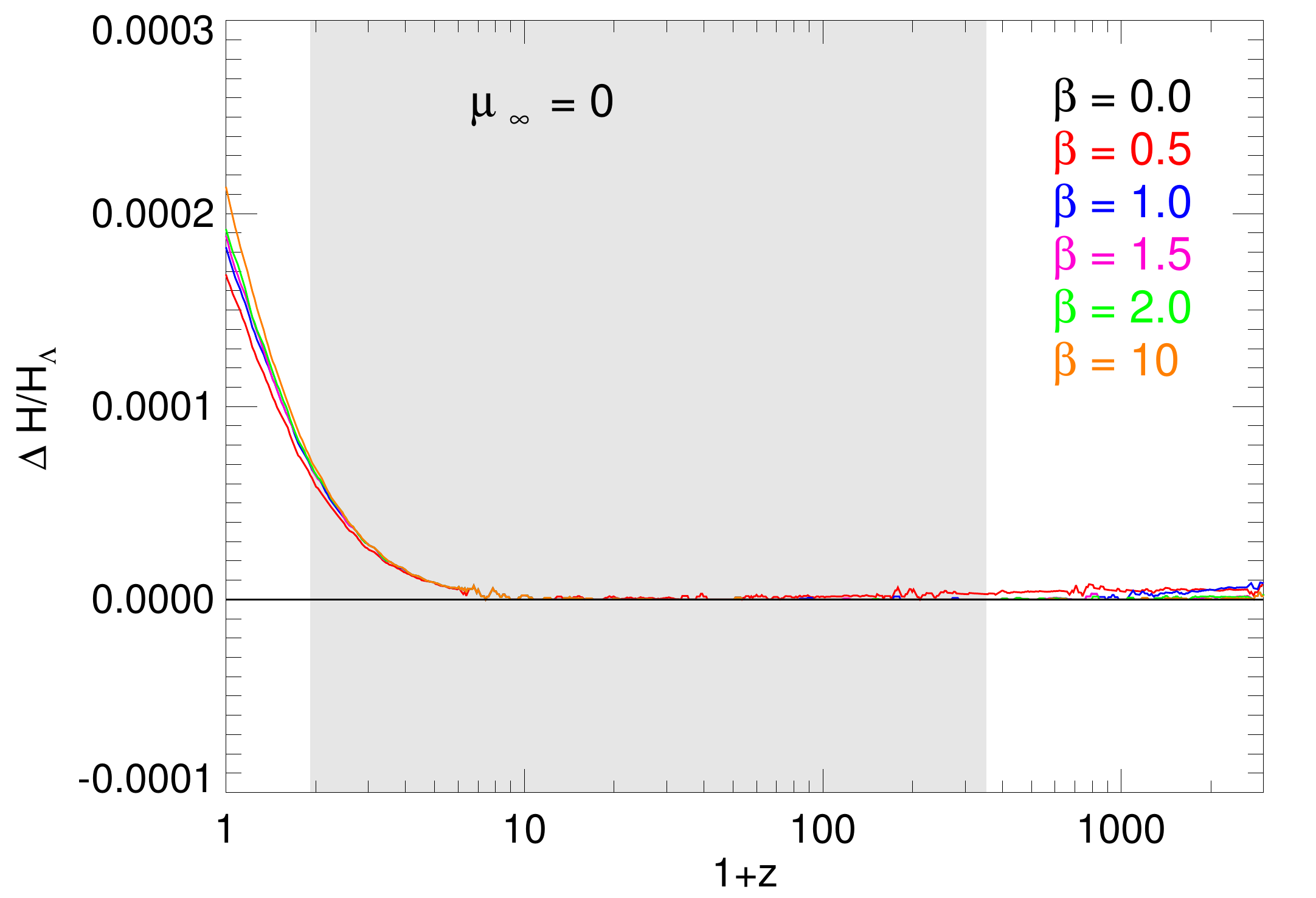}
\caption{{\em Left}: The effective coupling $\beta _{\rm eff}$ for a series of initially asymmetric McDE cosmologies ($\mu _{\infty }\neq 0$) with different coupling values. All the models evolve towards the symmetric attractor during matter domination. {\em Right}: The relative difference of the Hubble function $H(z)$ over the $\Lambda $CDM Hubble function $H_{\Lambda }$ for a series of initially symmetric McDE models. The expansion history of symmetric McDE models is almost indistinguishable from $\Lambda $CDM even for couplings as large as $\beta = 10$.}
\label{fig:1}
\end{figure}

In fact, previous investigations \cite{Brookfield_vandeBruck_Hall_2008,Baldi_2012a} have shown that  during matter domination the symmetric state $\mu = 0$ is an attractor of the system, such that even for initially asymmetric states the system moves towards symmetry at the beginning of matter domination, and remains in such state until the emergence of DE. Therefore, the existence of a symmetric attractor for the McDE background system dynamically realises  the conditions for the effective screening described by Eq.~\ref{effective_coupling}. This basic feature of the McDE model ensures that even large values of the coupling have a very mild impact on the background evolution of the Universe, thereby evading most of the
present observational constraints. This is well explained in the two panels of Fig.~\ref{fig:1}, where the dynamical suppression of the effective coupling during matter domination for initially asymmetric states with an initial asymmetry of $\mu _{\infty } = \pm 0.5$ is shown in the left plot, while the deviation from the Hubble expansion history of an uncoupled system (i.e. identical to $\Lambda $CDM) at the end of matter domination is shown in the right plot for an initially symmetric state. 
The latter is a consequence of the dynamical evolution of the scalar field that after being frozen during matter domination in the minimum of the effective potential defined by:
\begin{equation}
\frac{d V_{\rm eff}}{d\phi }\equiv \frac{dV}{d\phi} -C\rho _{+} + C\rho _{-}\,,
\end{equation}
starts moving again as soon as DE takes the lead of the cosmic budget. McDE models therefore also naturally provide a time-dependent effective coupling (see e.g. \cite{Baldi_2011a}) without imposing {\em a priori} any specific form for the coupling evolution.
The right plot of Fig.~\ref{fig:1} clearly shows that in any case, even for very large values of the coupling, the deviation of the Hubble function from the $\Lambda $CDM case never exceeds a fraction of a percent, thereby being undetectable with the present level of observational precision.

\subsection{Linear Perturbations}
As discussed in \cite{Baldi_2012a}, the evolution of linear density perturbations in McDE cosmologies is described by the equations:
\begin{eqnarray}
\label{gf_plus}
\ddot{\delta }_{+} &=& -2H\left[ 1 - \beta \frac{\dot{\phi }}{H\sqrt{6}}\right] \dot{\delta }_{+} + 4\pi G \left[ \rho _{-}\delta _{-} \Gamma_{R} + \rho _{+}\delta _{+}\Gamma_{A}\right] \,, \\
\label{gf_minus}
\ddot{\delta }_{-} &=& -2H\left[ 1 + \beta \frac{\dot{\phi }}{H\sqrt{6}}\right] \dot{\delta }_{-} + 4\pi G \left[ \rho _{-}\delta _{-} \Gamma _{A} + \rho _{+}\delta _{+}\Gamma_{R}\right]\,,
\end{eqnarray}
where $\delta _{\pm } = \delta \rho _{\pm }/\rho _{\pm }$ is the density contrast of the two different CDM species.

The $\Gamma $ factors in Eqs.~\ref{gf_plus}-\ref{gf_minus} are defined as
\begin{equation}
\label{def_gamma}
\Gamma _{A} \equiv 1 + \frac{4}{3}\beta ^{2}\,, \quad \Gamma _{R}\equiv 1 - \frac{4}{3}\beta ^{2} \,,
\end{equation}
and encode the attractive ($\Gamma _{A}$) or repulsive ($\Gamma _{R}$) corrections to standard gravity due to the
long-range fifth-force mediated by the DE scalar field, 
while the different signs of the extra-friction term (second term in the first squared brackets on the right-hand side) are a consequence of the of the opposite mass evolution of the two CDM particle types.

Since a dimensionless coupling $\beta $ determines a correction to standard gravity in Eqs.~\ref{gf_plus}-\ref{gf_minus} of order $\beta ^{2}$, a coupling of order unity or larger might provide a scalar fifth-force with strength comparable or even larger than standard gravity, while present constraints on standard cDE models with a single CDM species based on different types of observables provide an upper limit of $\beta \lesssim 0.08-0.15$ \cite{Bean_etal_2008,Xia_2009,Baldi_Viel_2010,Pettorino_etal_2012} corresponding to a few percent correction to standard gravity. 
More specifically, a coupling of $\beta = \sqrt{3}/2 \approx 0.87$ 
would induce a scalar force with the same strength as gravity, thereby resulting in the absence of any force for repulsive corrections ($\Gamma _{R} = 0$), and in a force twice as strong as gravity for attractive corrections ($\Gamma _{A} = 2$), while a coupling of $\beta =\sqrt{3/2}\approx 1.22$ would imply an attractive total force with three times the strength of gravity for attractive corrections ($\Gamma _{A} = 3$), and a repulsive force with gravitational strength for repulsive corrections ($\Gamma _{R} = -1$). 

\begin{figure}
\begin{center}
\includegraphics[height=55mm]{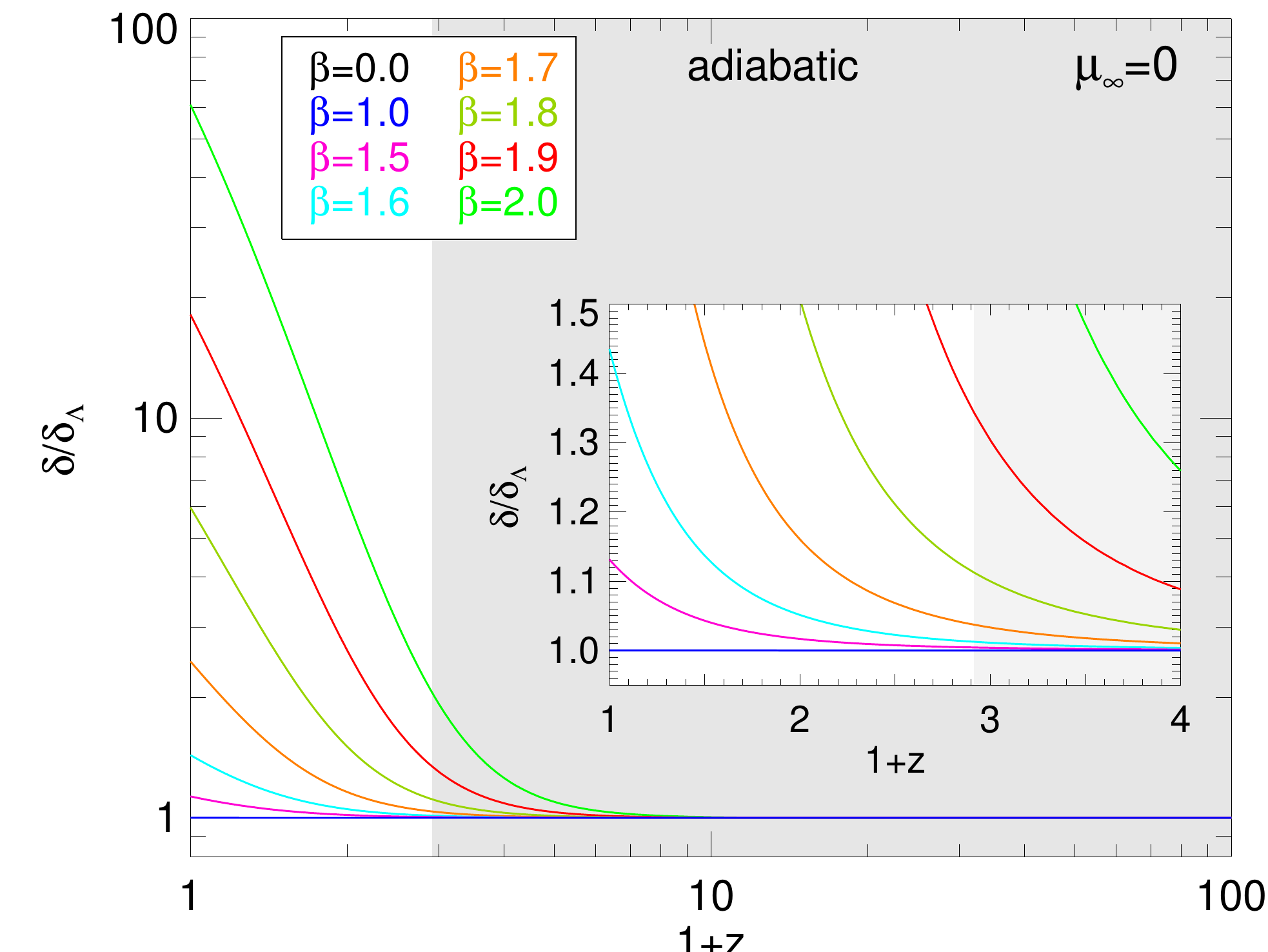}
\end{center}
\caption{The ratio of the growth factor of CDM density perturbations to the $\Lambda $CDM case for a series of McDE models with couplings between $\beta = 1$ and $\beta = 2$ and with adiabatic initial conditions.}
\label{fig:2}
\end{figure}

If we restrict to the case of initially symmetric models and of initially adiabatic density perturbations of the two CDM species (for a more general discussion of asymmetric and non-adiabatic models see \cite{Baldi_2012a}), i.e.
to the case:
\begin{eqnarray}
\mu _{\infty } &=& 0 \,, \\
\delta _{+}(z_{\infty }) &=& \delta _{-}(z_{\infty })\,,
\end{eqnarray}
Eqs.~\ref{gf_plus}-\ref{gf_minus} reduce to:
\begin{eqnarray}
\label{gf_plus_adiab}
\ddot{\delta }_{+} &=& -2H\left[ 1 - \beta \frac{\dot{\phi }}{H\sqrt{6}}\right] \dot{\delta }_{+} + 8\pi G\rho _{+}\delta _{+}\,, \\
\label{gf_minus_adiab}
\ddot{\delta }_{-} &=& -2H\left[ 1 + \beta \frac{\dot{\phi }}{H\sqrt{6}}\right] \dot{\delta }_{-} + 8\pi G\rho _{-}\delta _{-}\,.
\end{eqnarray}
For this choice of initial conditions, we investigate the dynamics of the total linear density perturbations defined as:
\begin{equation}
\delta _{\rm CDM} \equiv \frac{\Omega _{+}\delta _{+}}{\Omega _{\rm CDM}} + \frac{\Omega _{-}\delta _{-}}{\Omega _{\rm CDM}} \,
\end{equation}
by numerically solving Eqs.~\ref{gf_plus_adiab}-\ref{gf_minus_adiab} for different values of the coupling $\beta $. The ratio of the linear growth
factor obtained in this way to the standard $\Lambda $CDM growth factor is shown in Fig.~\ref{fig:2}. As the plot clearly shows, while at the background level
the effective screening of the coupling due to the attractor symmetric state of the system suppresses almost completely the interaction and allows couplings as large
as $\beta = 10$ to be observationally viable, the evolution of linear density perturbations poses much tighter constraints on the coupling, as the linear growth factor significantly deviates from the $\Lambda $CDM case at low redshifts even for coupling values as small as $\beta =2$. More specifically, as one can see in the plot, couplings larger than $\beta = 1.6$ (cyan curve) determine an overall enhancement of the amplitude of linear density perturbations at $z=0$ exceeding $40 \%$, while a coupling $\beta = 1.5$ induces just a $\sim 13\%$ enhancement,  and a coupling $\beta = 1$ does not produce any appreciable deviation in the growth. This clearly shows that the screening effect of the matter-dominated attractor of the McDE system is broken at the level of linear density perturbations, which potentially allow to distinguish a McDE model from an uncoupled cosmology, even for coupling values that would be completely undetectable based on pure background observations. Nevertheless, the effect of enhanced growth at low redshifts appears to be very strongly dependent on the coupling value, such that a coupling of order unity does not show any enhancement at all,  thereby resulting still indistinguishable from $\Lambda $CDM. Couplings of order unity therefore cannot be ruled out even at the linear level in McDE models, and this is of course equally true for even smaller couplings $\beta \lesssim 1$. 

\section{Non-linear evolution}
\label{nonlinear}

Since the background and linear perturbations evolution of McDE scenarios still leave a large portion of the parameter space as fully viable, the next step in the investigation of McDE is the exploration of the nonlinear regime of structure formation, by means of dedicated N-body simulations. Such analysis has been performed by \cite{Baldi_2012c} making use of the N-body code {\small C-GADGET} \cite{Baldi_etal_2010}. 
Such code was initially developed as a modified version of the widely used Tree-PM N-body code {\small GADGET-3} \cite{gadget} for standard cDE models, and has been widely employed in the last years to investigate different realisations of interacting DE cosmologies with constant couplings \cite{Baldi_etal_2011a,Baldi_2011b,Baldi_Pettorino_2011}, variable couplings \cite{Baldi_2011a,Baldi_Lee_Maccio_2011,Lee_Baldi_2011}, and with "bouncing" DE potentials \cite{Baldi_2011c,CoDECS}. Since the same code can be directly applied also to the case of multiple CDM families with individual couplings, \cite{Baldi_2012c} have run some intermediate-resolution simulations of McDE cosmologies for six different values of the coupling
($\beta = 0\,, 1/2\,, \sqrt{3}/2\,, 1\,, \sqrt{3/2}\,, 3/2$)
with the aim to highlight the main qualitative features of McDE models in the nonlinear regime. 
Such simulations
have followed the evolution of $2\times 256^{3}$ CDM particles in a periodic cosmological box of $100$ Mpc$/h$ aside, with a mass resolution of $m_{\pm } = 2.24\times 10^{9}$ M$_{\odot }/h$, and a gravitational softening  
$\epsilon _{g} = 10$ kpc$/h$. 
Initial conditions have been
generated by rescaling the amplitude of the linear matter power spectrum obtained with the Boltzmann code {\small CAMB} \cite{camb} for a standard $\Lambda $CDM universe with cosmological parameters in accordance with WMAP7 results \cite{wmap7} between last scattering ($z\approx 1100$) and the starting redshift of the simulations by applying the specific growth factor computed for each model by numerically integrating Eqs.~(\ref{gf_plus},\ref{gf_minus}).

\begin{figure*}
\begin{center}
\begin{minipage}{65mm}
\begin{center}
{\large $\beta = 0$}
\end{center}
\end{minipage}
\begin{minipage}{65mm}
\begin{center}
{\large $\beta = 1/2$}
\end{center}
\end{minipage}\\
\begin{minipage}{65mm}
\includegraphics[width=\linewidth]{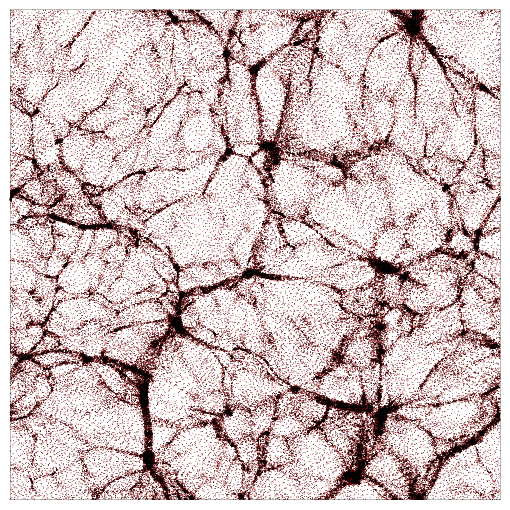}
\end{minipage}
\begin{minipage}{65mm}
\includegraphics[width=\linewidth]{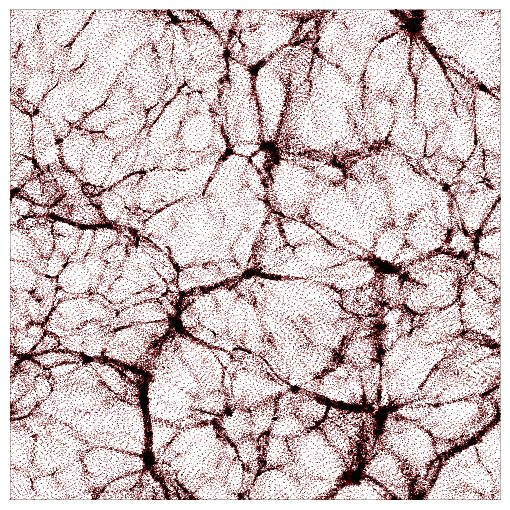}
\end{minipage}\\
\begin{minipage}{65mm}
\begin{center}
{\large $\beta = \sqrt{3}/2$}
\end{center}
\end{minipage}
\begin{minipage}{65mm}
\begin{center}
{\large $\beta = 1$}
\end{center}
\end{minipage}\\
\begin{minipage}{65mm}
\includegraphics[width=\linewidth]{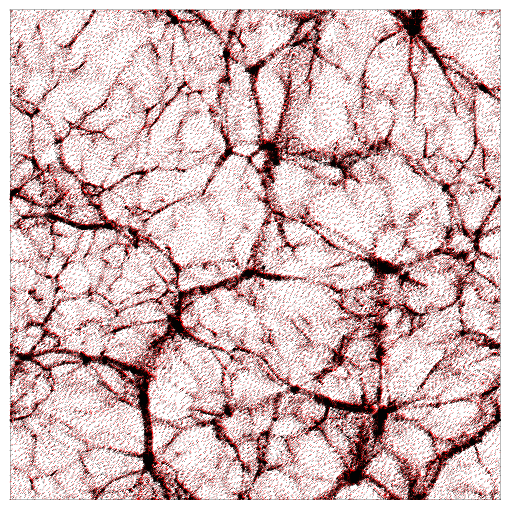}
\end{minipage}
\begin{minipage}{65mm}
\includegraphics[width=\linewidth]{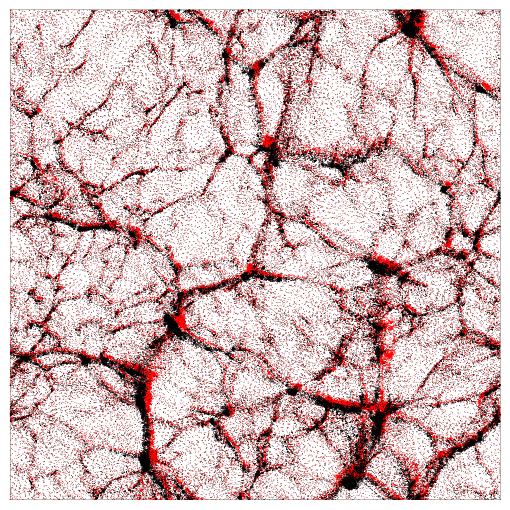}
\end{minipage}\\
\begin{minipage}{65mm}
\begin{center}
{\large $\beta = \sqrt{3/2}$}
\end{center}
\end{minipage}
\begin{minipage}{65mm}
\begin{center}
{\large $\beta = 3/2$}
\end{center}
\end{minipage}\\
\begin{minipage}{65mm}
\includegraphics[width=\linewidth]{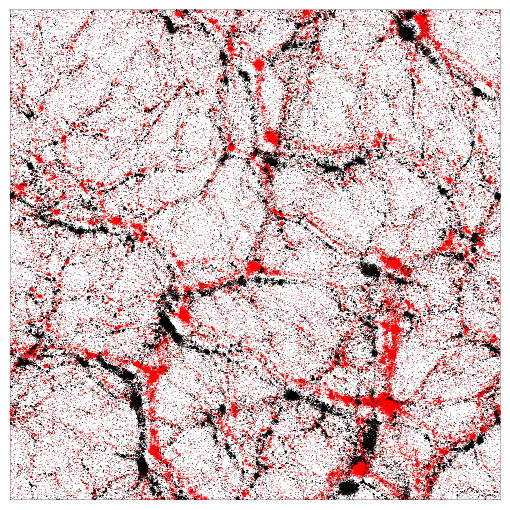}
\end{minipage}
\begin{minipage}{65mm}
\includegraphics[width=\linewidth]{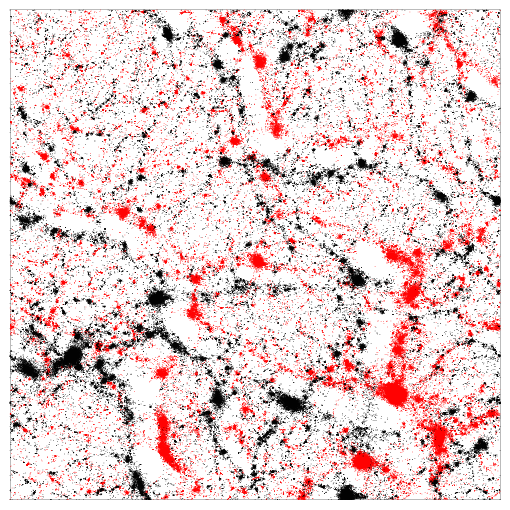}
\end{minipage}
\end{center}
\caption{The distribution of the positively- and negatively-coupled CDM particles (red and black points, respectively) in a slice of $100\times 100 \times 2$ Mpc$/h$ for different values of the coupling $\beta $.}
\label{fig:slices}
\end{figure*}

Figure~\ref{fig:slices} shows the projected position
of particles of the positively- and negatively-coupled CDM species (as red and black points, respectively) in a $2$ Mpc$/h$-thick slice of the cosmological box
of the different simulations at $z=0$. The figure clearly shows that in the absence of coupling ($\beta =0$, left panel of the upper row) the red points (always plotted first such that they might be covered by the black ones) are barely visible (except in voids) as their spatial distribution matches that of the negatively-coupled particles represented by the black points: this is the expected behaviour for the case of two different particle species obeying the same gravitational interaction law. It is already interesting to notice that the situation does not significantly change for the case of a coupling $\beta = 1/2$, that would be ruled out at more than 5 $\sigma $ C.L. \cite{Bean_etal_2008,Xia_2009,Baldi_Viel_2010} for the case of a single coupled CDM species, for which the qualitative shape of the large-scale structures appears indistinguishable
from the uncoupled case. This result already shows that a scalar fifth-force with one third the strength of standard gravity
does not affect in a significant way  the formation of large-scale cosmic structures even in the nonlinear regime probed with full N-body simulations. More detailed investigations of the impact of such a coupling value on the innermost structure of highly nonlinear collapsed objects is presently ongoing. 

\normalsize

Moving to progressively larger couplings, some clear and characteristic effects on the shape of large-scale structures start to appear, with the two different CDM particle species
being no longer identically distributed in space, even though for a coupling of $\beta = \sqrt{3}/2$ (corresponding to a scalar fifth-force as strong as standard gravity)  the overall effect is still relatively mild. For even larger coupling values, the two CDM particle species start to show a clear shift in their distributions, due to the repulsion of particles of opposite type as a consequence of their opposite coupling to DE. This effect becomes progressively more dramatic for increasing coupling values, leading to the distinct formation of two independent and almost equally-shaped cosmic networks. The development of such "mirror" structures represents a characteristic footprint of McDE models in the nonlinear regime of structure formation, possibly allowing to constrain the model beyond the relatively loose bounds allowed by background and linear perturbation analyses.

Even such a qualitative inspection of the outcomes of these six simulations already shows quite clearly how the nonlinear regime of structure formation can highlight characteristic observational features of McDE scenarios that were proven to be almost indistinguishable from the standard $\Lambda $CDM cosmology at the level of background and linear perturbations observables, thereby offering an additional possibility to test and constrain the model. Nonetheless, for coupling values smaller than $\beta = \sqrt{3}/2$ such effects appear still relatively mild. 

\begin{figure*}
\begin{center}
\includegraphics[scale=0.40]{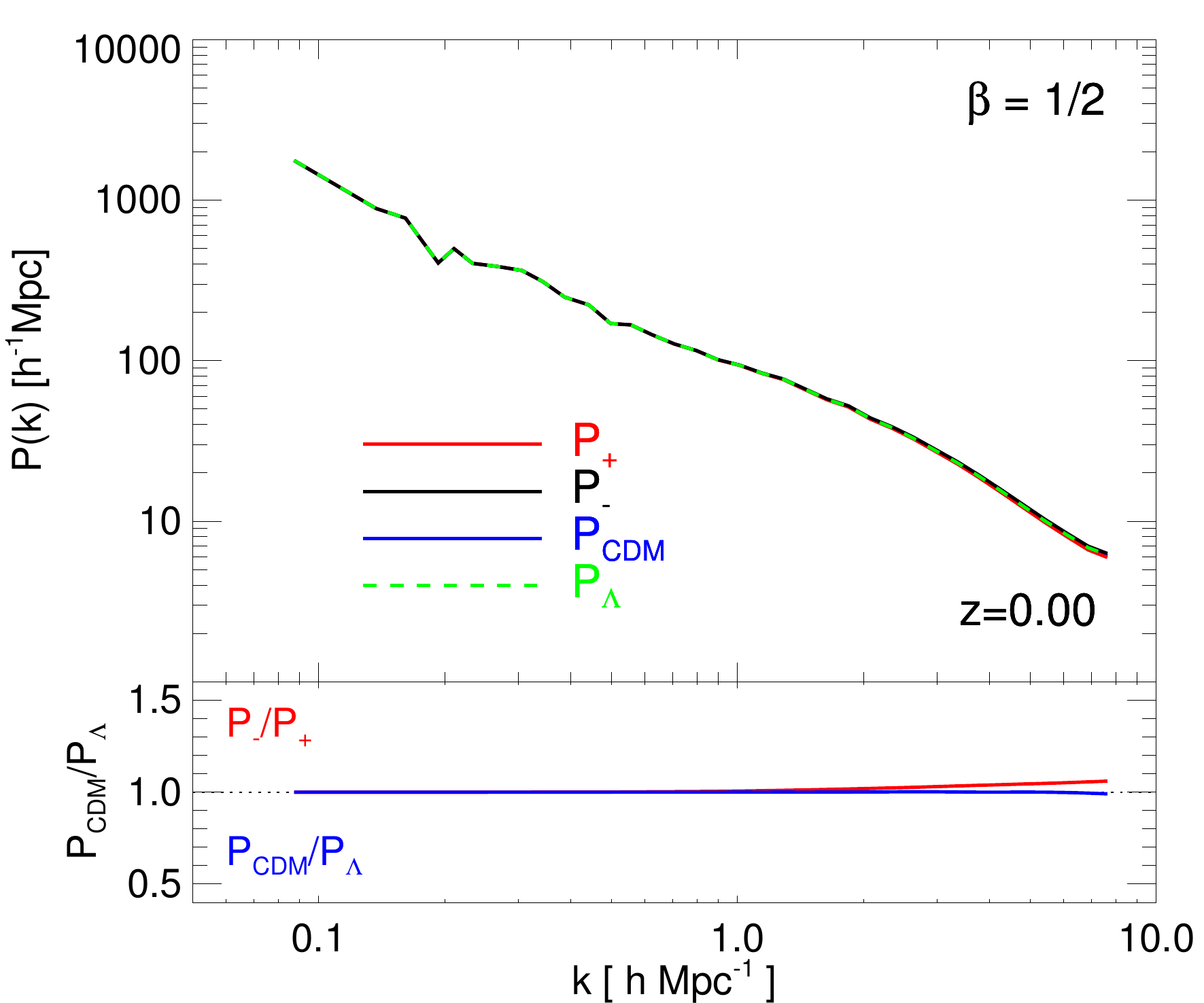}
\includegraphics[scale=0.40]{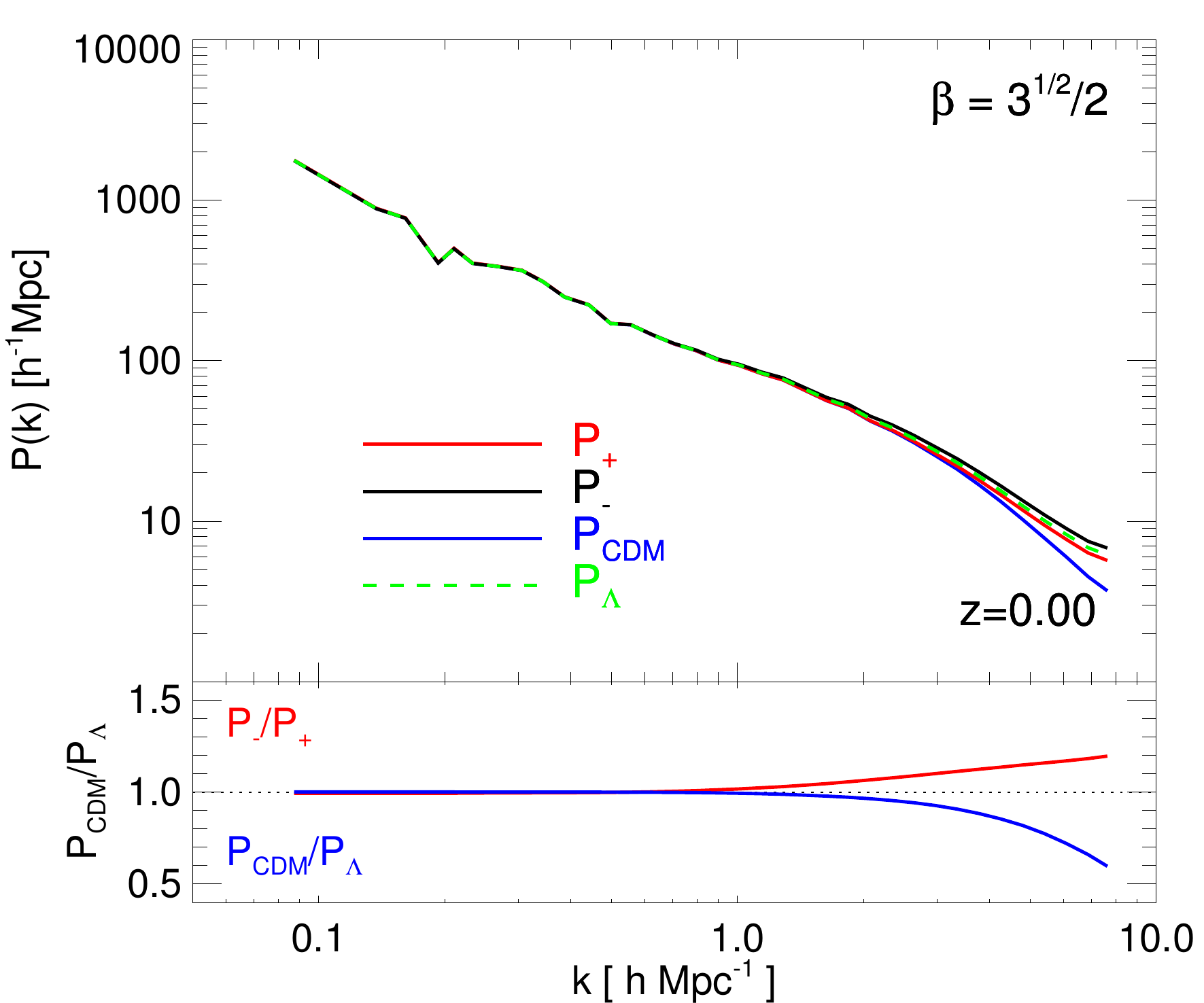}\\
\includegraphics[scale=0.40]{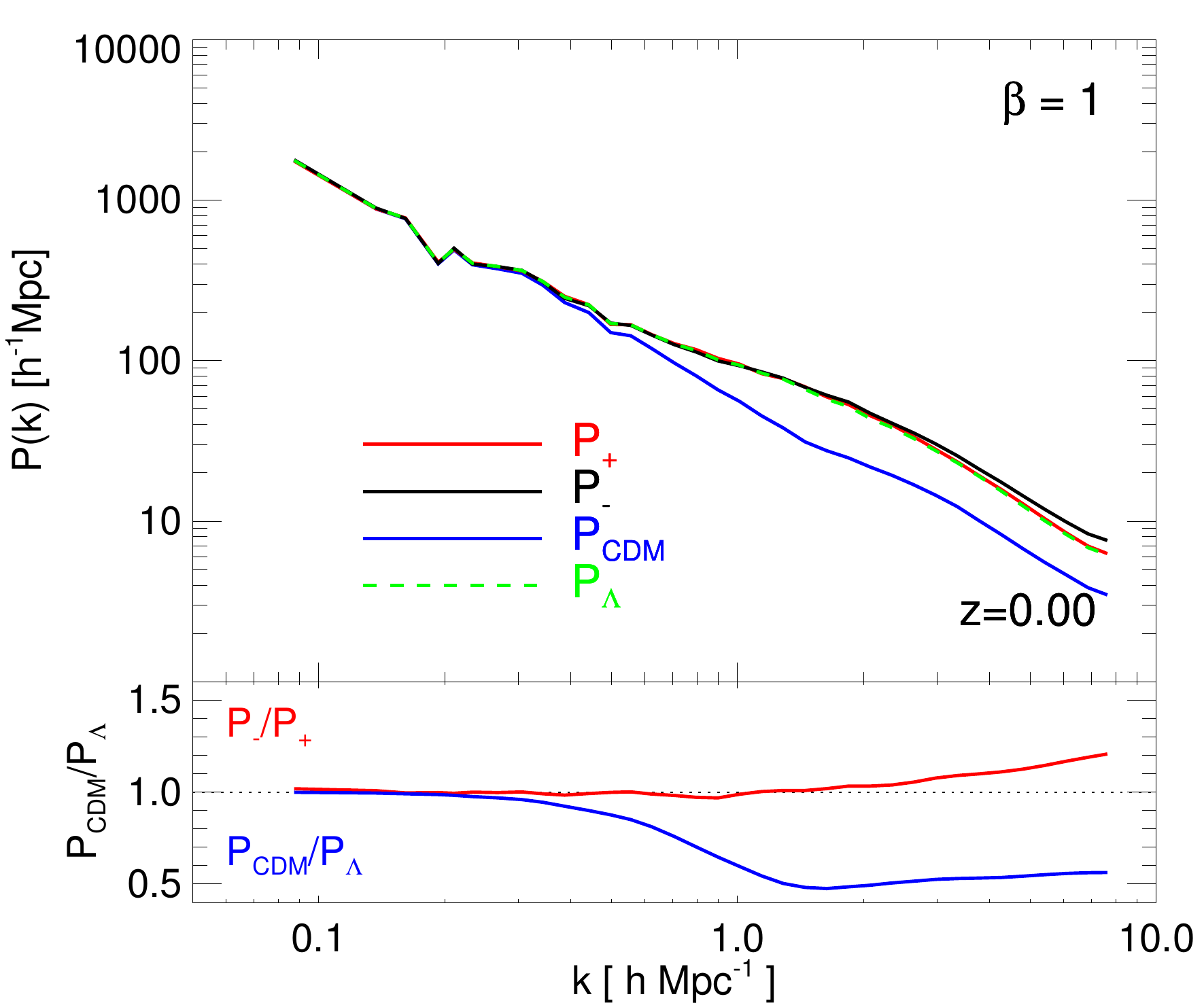}
\includegraphics[scale=0.40]{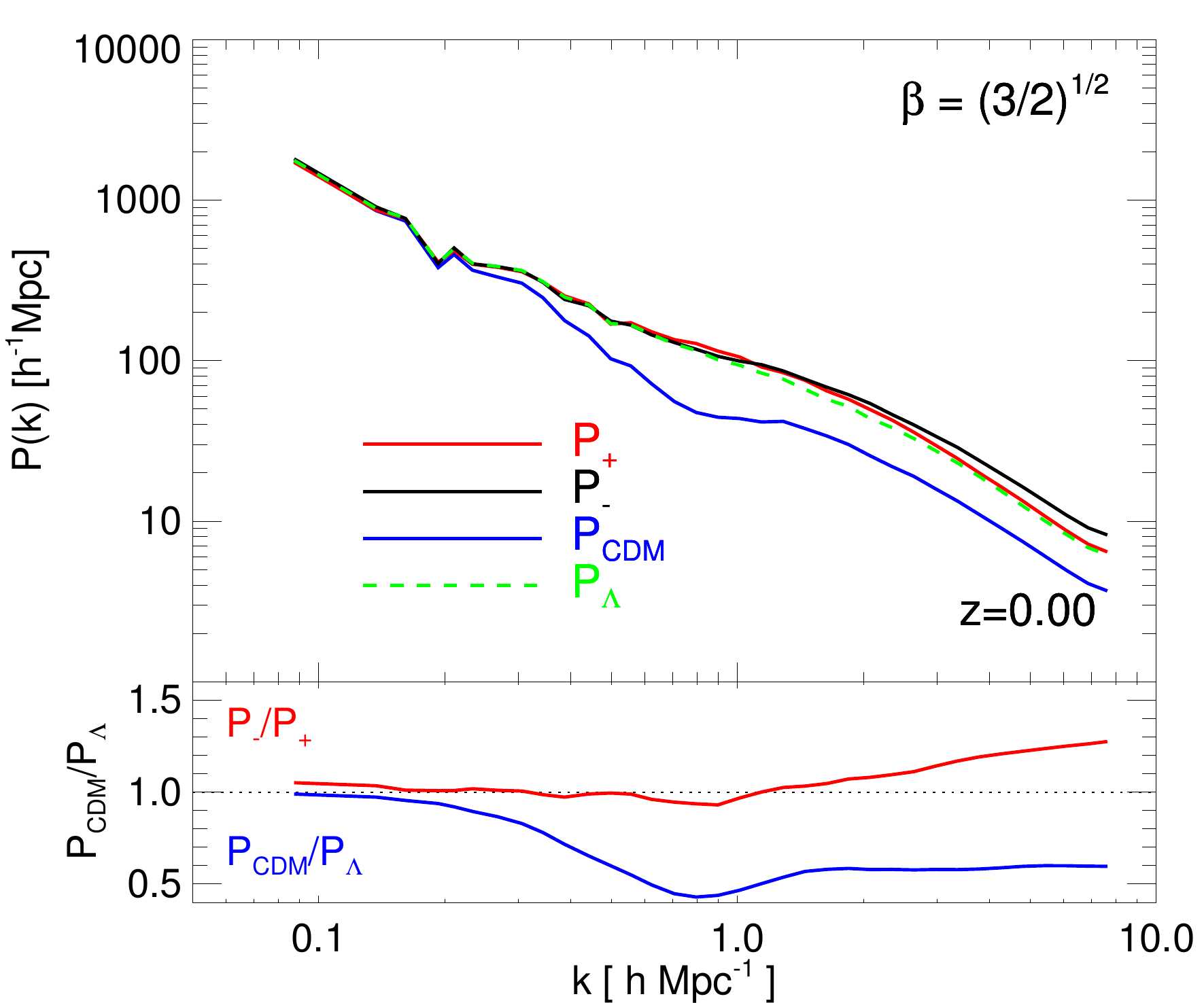}\\
\includegraphics[scale=0.40]{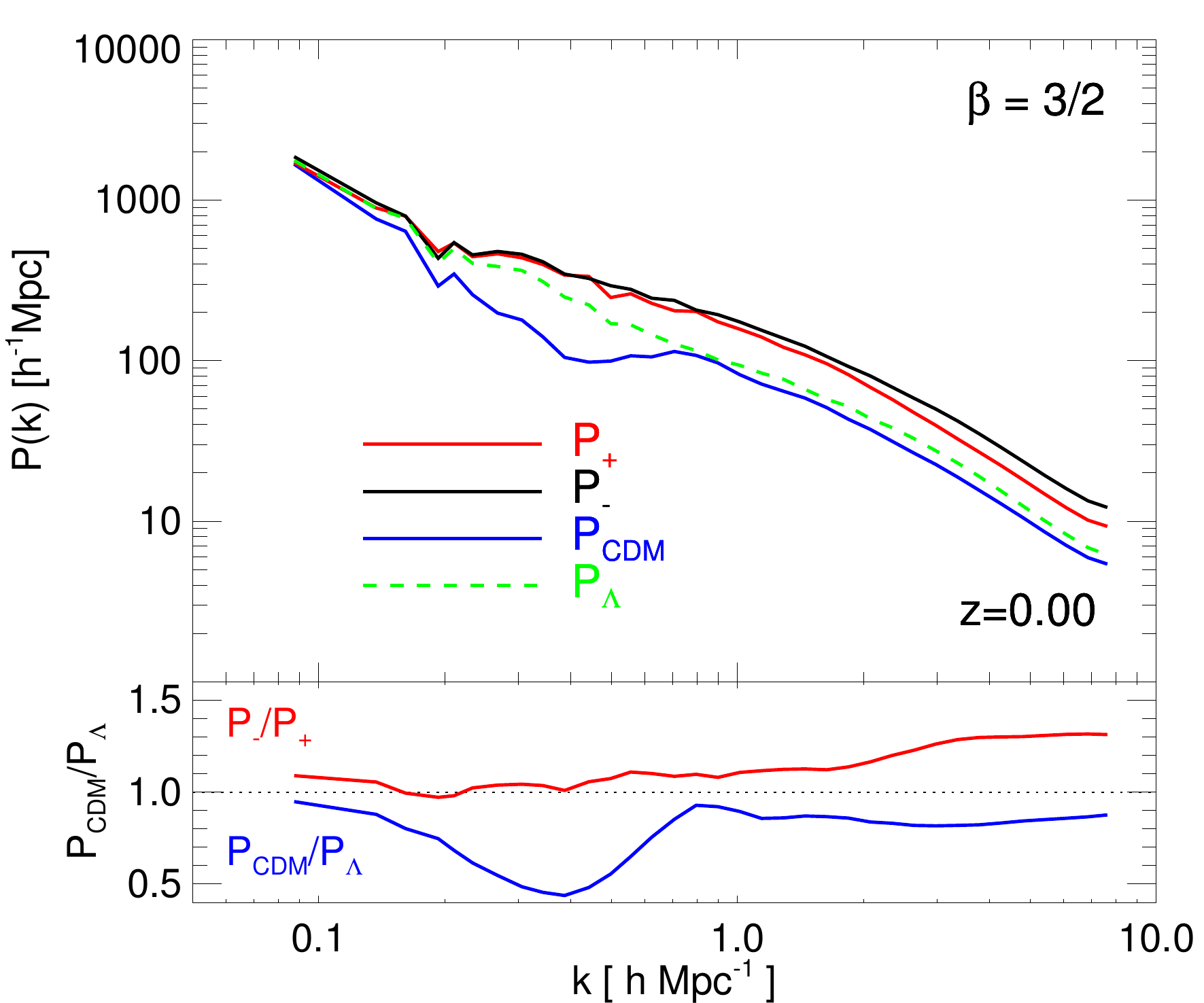}
\end{center}
\caption{The upper plot of each panel shows the power spectrum of the different CDM species $P_{+}(k)$ and $P_{-}(k)$ and of the total CDM component $P_{\rm CDM}(k)$, plotted as red, black, and blue solid lines, respectively, and the matter power spectrum of the standard $\Lambda $CDM cosmology as a green dashed line, all at $z=0$. The bottom plots show in red the ratio of the power spectra of the two different CDM components $P_{-}(k)/P_{+}(k)$ and in blue the ratio of the total CDM power spectrum to the $\Lambda $CDM reference case, $P_{\rm CDM}/P_{\Lambda }$.}
\label{fig:powerspectra}
\end{figure*}

In order to better quantify the impact of McDE models on structure formation, going beyond the rather qualitative visual inspection of simulated large-scale structures that we discussed above, one can compute the matter power spectrum associated with the distribution of the two different CDM species in the different simulations outputs at $z=0$, and compare the total matter power spectrum with the one obtained in the absence of any coupling. This is done in Fig.~\ref{fig:powerspectra}, where the separate matter power spectra of the two different CDM species are plotted as red and black solid curves ($P_{+}(k)$ and $P_{-}(k)$, respectively), while the total CDM power spectrum is plotted in blue, and the reference $\Lambda $CDM power is also included for reference as a dashed green curve. The bottom plot of each panel of Fig.~\ref{fig:powerspectra} shows the ratio of the power spectra of the two different CDM species $P_{-}(k)/P_{+}(k)$ in red, while the ratio of the total CDM power to the standard $\Lambda $CDM expectation is displayed in blue.

As the different plots clearly show, a coupling $\beta = 1/2$ has almost no impact on the power spectra at $z=0$ over the whole range of scales probed by our simulations. This result quantitatively confirms the qualitative conclusion obtained by visually inspecting Fig.~\ref{fig:slices}. However, for larger coupling values, already starting with $\beta = \sqrt{3}/2$, some effects become apparent both on the relative shape of the power spectra of the two different CDM species and on the relative behaviour of the total CDM power as compared to the standard $\Lambda $CDM case. 
In particular, for $\beta = \sqrt{3}/2$, corresponding to a scalar fifth-force with gravitational strength, 
a suppression of the total CDM power spectrum is visible 
for $k\gtrsim 1\, h/$Mpc, and further increases for even smaller  scales, reaching
a maximum suppression of $\sim 40\%$ at the Nyquist frequency of the box ($k_{\rm Ny} \sim 8\, h/$Mpc), while for $k\lesssim 1\, h/$Mpc the effect is practically absent and all the different power spectra are still indistinguishable. This is a very interesting result since it shows a characteristic observational feature of a long-range scalar force with gravitational strength that appears only at scales that will be efficiently probed by upcoming weak lensing surveys, while  larger scales as well as the background evolution of the universe remain completely unaffected.

These effects become progressively larger when moving to larger coupling values, with a suppression of the total CDM power that can reach a level of about $40-50\%$ already at $k\sim 2\,h/$Mpc. The plateau in the ratio of the total CDM power spectrum over the $\Lambda $CDM expectation that appears in the plots for
$\beta \geq 1$ at small scales, leaving a maximum in the power spectrum suppression at intermediate scales, can be explained in terms of the progressive fragmentation -- starting from the smallest scales that are the first to grow nonlinear -- of collapsed object made of a mixture of the two CDM particle types into smaller objects composed primarily by one single particle species. Such fragmentation is triggered by local deviations from perfect spherical symmetry that makes the repulsive fifth-force between opposite types of CDM particles to pull them apart (see \cite{Baldi_2012c} for a more detailed discussion on the fragmentation process in McDE).

\normalsize

\section{Conclusions}
\label{concl}

In the present paper, I have presented a general overview of a new class of cosmological scenarios characterised by the interaction of a Dark Energy scalar field with two different species of Cold Dark Matter particles, through individual couplings with equal absolute value but opposite sign. This determines the existence of both attractive and repulsive fifth-forces between CDM particles, which have a very significant impact on the evolution of structure formation processes. Such particular model
that has been recently proposed in the literature and termed the "Multi-coupled Dark Energy" scenario, does not require additional free parameters with respect to standard coupled Quintessence models, but has the additional appealing feature of providing a dynamical screening of the coupling during matter domination, thanks to the presence of a stable attractor of the background dynamical system characterised by a vanishing effective coupling. As a consequence of this particular behaviour, Multi-coupled Dark Energy models allow for much larger values of the interaction strength as compared to standard coupled Quintessence models, for which present bounds constrain the coupling parameter to $\beta \lesssim 0.1$. For Multi-coupled Dark Energy models, instead, coupling values as large as $\beta =10$ appear almost indistinguishable from the standard $\Lambda $CDM background evolution, as the coupling becomes active only at the onset of Dark Energy domination.\\

The analysis of the evolution of linear density perturbations in Multi-coupled Dark Energy scenarios, shows that such effective screening that suppresses the coupling at the background level is to some extent broken for linear perturbations, due to the instability associated to the repulsive fifth-force experienced by density perturbations in the two different CDM species. Such instability becomes apparent in a timescale shorter than the Hubble time only for relatively large values of the coupling $\beta $, but nevertheless allows to significantly restrict the viable range of parameters to coupling values $\beta \lesssim 1.5$, thereby providing a much stronger constraining power as compared to the background evolution. Nonetheless, couplings of order unity, corresponding to a fifth-force with the same strength of standard gravity, still appear completely indistinguishable from the  $\Lambda $CDM predictions even at the level of linear density perturbations. \\

The next step in the investigation of Multi-coupled Dark Energy models is then clearly the exploration of the nonlinear regime of structure formation, where the direct effects of attractive and repulsive fifth-forces with a strength comparable to ordinary gravity are expected to have a significant impact. Such analysis, however, requires to adapt standard cosmological N-body algorithms by including all the additional physical effects associated with the Dark Energy interaction and with the existence of two distinct Cold Dark Matter species. This is allowed by the code {\small C-GADGET} that was originally developed for standard Coupled Quintessence models and that can be easily applied also to the case of Multi-coupled Dark Energy scenarios. By using such code, it has been possible to test the evolution of nonlinear structures in the context of six different Multi-coupled Dark Energy cosmologies with couplings ranging from $\beta =0$ to $\beta = 3/2$, and to highlight the most prominent features arising in the different models. Interestingly, this analysis has shown that coupling values as large as $\beta =\sqrt{3}/2$ still have a very little impact on the large-scale shape of cosmic structures, and on the total Cold Dar Matter power spectrum, even though a slight suppression of power starts to appear at the smallest scales ($k\gtrsim 1\, h/$Mpc). This is associated with ongoing processes of halo fragmentation that occur as a consequence of the repulsive fifth-force. For larger values of the coupling, the effect of the interaction on both the large-scale distribution of structures and the 
resulting matter power spectrum becomes more prominent and might possibly allow a direct constraint on the model using galaxy correlation functions or weak lensing observables. In particular, already a coupling of order unity, that was found to be completely indistinguishable from the uncoupled case at the background and linear perturbations level, shows a significative suppression (of about a factor $2$) of the total Cold Dark Matter power at nonlinear scales, whereas the simulations results confirm the negligible impact at linear scales already found through a linear perturbations analysis. \\

To conclude, Multi-coupled Dark Energy models have been investigated over a wide range of possible observational regimes, from their impact on the background expansion history of the Universe, to the effects on nonlinear structure formation. More detailed studies on the highly nonlinear regime characterising the internal structure of collapsed objects are presently ongoing, and will be required to further constrain the scenario. However, the analysis carried out so far already allowed to significantly restrict the viable parameter space of the model, and to highlight possible characteristic observational footprints for those parameter values that appear still compatible with present observational bounds. As a final result of this investigation, it is interesting to stress that scalar fifth-forces of gravitational strength in the dark sector do not seem to be ruled out yet under the not so unlikely assumption of a hidden internal symmetry of the Cold Dark Matter sector.

\section*{Acknowledgements}
I am deeply thankful to the organisers of the Corfu Summer Institute 2012 "School and Workshops on Elementary Particle Physics and Gravity" for the opportunity to present these results in such a pleasant location and constructive atmosphere. My talk as well as this proceeding contribution are based on the two previous publications \cite{Baldi_2012a,Baldi_2012c}. The present work has been supported by 
the DFG Cluster of Excellence "Origin and Structure of the Universe'', by
the TRR33 Transregio Collaborative Research Network on the "Dark%
Universe'', and by the Marie Curie
Intra European Fellowship "SIDUN" within the 7th
Framework Programme of the European Community.

\bibliographystyle{unsrt}
\bibliography{baldi_bibliography}

\end{document}